\documentclass[12pt]{iopart}
\usepackage{latexsym}
\usepackage{amscd}
\begin{document}
\jl{1} 
 \def\lambdabar{\protect\@lambdabar}
\def\@lambdabar{%
\relax
\bgroup
\def\@tempa{\hbox{\raise.73\ht0
\hbox to0pt{\kern.25\wd0\vrule width.5\wd0
height.1pt depth.1pt\hss}\box0}}%
\mathchoice{\setbox0\hbox{$\displaystyle\lambda$}\@tempa}%
{\setbox0\hbox{$\textstyle\lambda$}\@tempa}%
{\setbox0\hbox{$\scriptstyle\lambda$}\@tempa}%
{\setbox0\hbox{$\scriptscriptstyle\lambda$}\@tempa}%
\egroup}

\def\bbox#1{%
\relax\ifmmode
\mathchoice
{{\hbox{\boldmath$\displaystyle#1$}}}%
{{\hbox{\boldmath$\textstyle#1$}}}%
{{\hbox{\boldmath$\scriptstyle#1$}}}%
{{\hbox{\boldmath$\scriptscriptstyle#1$}}}%
\else
\mbox{#1}%
\fi
}
\def\msf{\hbox{{\sf M}}}
\def\psf{\hbox{{\sf P}}}
\def\Nsf{\hbox{{\sf N}}}
\def\Tsf{\hbox{{\sf T}}}
\def\Asf{\hbox{{\sf A}}}
\def\Bsf{\hbox{{\sf B}}}

\def\msfsim{\msf_{{\scriptstyle \rm{(sym)}}}}
\newcommand{\mcsim}{ {\sf M}_{ {\scriptstyle \rm{(sym)} } i_1\dots i_n}}
\newcommand{\mcs}{ {\sf M}_{ {\scriptstyle \rm {(sym)} } i_1i_2i_3}}

\newcommand{\beqan}{\begin{eqnarray*}}
\newcommand{\eeqan}{\end{eqnarray*}}
\newcommand{\beqa}{\begin{eqnarray}}
\newcommand{\eeqa}{\end{eqnarray}}

 \newcommand{\suml}{\sum\limits}
\newcommand{\intl}{\int\limits}
\newcommand{\rvec}{\bbox{r}}
\newcommand{\xivec}{\bbox{\xi}}
\newcommand{\Avec}{\bbox{A}}
\newcommand{\Rvec}{\bbox{R}}
\newcommand{\Evec}{\bbox{E}}
\newcommand{\Bvec}{\bbox{B}}
\newcommand{\Svec}{\bbox{S}}
\newcommand{\avec}{\bbox{a}}
\newcommand{\nablav}{\bbox{\nabla}}
\newcommand{\nuvec}{\bbox{\nu}}
\newcommand{\bvec}{\bbox{\beta}}
\newcommand{\vvec}{\bbox{v}}
\newcommand{\jvec}{\bbox{j}}
\newcommand{\nvec}{\bbox{n}}
\newcommand{\pvec}{\bbox{p}}
\newcommand{\mvec}{\bbox{m}}
\newcommand{\evec}{\bi{e}}
\newcommand{\eps}{\varepsilon}
\newcommand{\la}{\lambda}
\newcommand{\rad}{\mbox{\footnotesize rad}}
\newcommand{\scr}{\scriptstyle}
\newcommand{\latens}{\bbox{\Lambda}}
\newcommand{\pitens}{\bbox{\Pi}}
\newcommand{\cm}{{\cal M}}
\newcommand{\cp}{{\cal P}}
\newcommand{\al}{\alpha}
\renewcommand{\d}{\partial}
\def\rmi{{\rm i}}
\def\rme{\hbox{\rm e}}
\def\rmd{\hbox{\rm d}}

\title{ Expressing  Maxwell's equations independently of the unit systems}
\author{C. Vrejoiu}
\address{Facultatea de Fizic\v{a}, Universitatea din Bucure\c{s}ti, 76900,
Bucure\c{s}ti, Rom\^{a}nia\\  E-mail : vrejoiu@shogun.fizica.unibuc.ro  }
\begin{abstract}
A procedure to teach Electrodynamics independently of unit systems 
is presented and compared with some of those given in physics literature.
\end{abstract}
\section{Introduction}
As it is remarked in \cite{leung}, "it is a well-known fact that one major 
hurdle for students in a class of electromagnetism is to get familiar with 
the adopted unit system, and to move from one unit system to another 
(e.g. SI to Gaussian)". As a student and, later, as an Electrodynamics 
professor, I have felt myself this hurdle. However, inspired by the Jackson's 
book \cite{jack}, I have adopted a procedure to teach electrodynamics 
independently of unit systems since 1985 and used it consistently in my 
lectures \cite{vre93}.
 Here is a summary, including some comments regarding results from 
 \cite{leung, jack} and some inconsistencies from \cite{vre93}.

\section{Writing the Maxwell's equations}
Before writing the Maxwell's equations  we have to define the physical 
system itself, namely the electromagnetic field (EMF), by the interactions 
with other known systems. The Lorentz force may be 
experienced for introducing the electric charge $q$ and the fundamental 
variables $\Evec$ and $\Bvec$ of the EMF. The electric field $\Evec$ 
is defined by the force acting on a charge at  rest  in the inertial 
system of the laboratory. The relation between force, charge and electric 
field, as is considered in all known unit systems \cite{jack} is
$$ \bbox{F}=q\Evec,$$
 such that the  introduction of an arbitrary proportional constant  in this relation 
 is not of practical interest.
Different unit systems are introduced when expressing the Lorentz force  
 acting on a moving charge $q$ \footnote{In \cite{vre93} the magnetic field is defined within  the old 
formalism of the {\it magnetic shells} or {\it sheets} and,  the 
constant $\al_0$ is introduced by the equation of the torque on an elementary shell  
 $\bbox{C}=(1/\al_0)I\bbox{S}\times\Bvec
=\bbox{m}\times\Bvec$ where the magnetic moment is $\bbox{m}=(1/\al_0)I\bbox{S}$.  
 From this torque one may deduce the Laplace's force 
 (up to a gradient) and from this force, the presence of the factor $1/\al_0$ in 
 the Lorentz force expression.}
 \begin{equation}\label{FL}
\;\;\;\;\;\;\;\;\;\;\;\;\;\;\;\;\;\;\;\;\;\;\;\;\;\;\;\;\;\;\;\;\;\;\;\;\;\;\;
\bbox{F}=q\Evec+\frac{q}{\al_0}\bbox{v}\times\Bvec,\end{equation}
Here,  the constant $\al_0$ depends on the unit system used.
\par The law of electric flux introduces a second constant $k_e$:
$$\oint\limits_{\d\mathcal{D}}\Evec\cdot\bbox{dS}=k_eQ(\mathcal{D})$$
or in the local form
$$\nablav\cdot\Evec=k_e\rho.$$
 Amp\`{e}re's law introduces a third constant $k_m$
$$\oint\limits_\Gamma\Bvec\cdot\bbox{dl}=
k_mI(\Sigma_\Gamma)\;\;\;\mbox{or}\;\;\; \nablav\times\Bvec=k_m\jvec.$$
 Maxwell's equation that generalizes the Amp\`{e}re's law is 
$$\nablav\times\Bvec=k_m\left(\jvec+\frac{1}{k_e}\frac{\d\Evec}{\d t}\right).$$
Let's adopt, for practical reasons, the following new notations:
$$k_e=\frac{1}{\tilde{\eps}_0},\;\;\;k_m=\frac{\tilde{\mu}_0}{\al_0}.$$
Here, the constants $\tilde{\eps}_0$ and $\tilde{\mu}_0$ are  
proportional constants, their values depending on the units used. 
 \footnote{In \cite{vre93}  the constant $\tilde{\mu}_0$ is introduced defining the magnetic scalar potential of an elementary magnetic shell 
$$\Psi(\rvec)=\frac{\tilde{\mu}_0}{4\pi}\frac{\bbox{m}\cdot\rvec}{r^3}.$$ Consequently,  
the factor $\tilde{\mu}_0$ occurs in the Amp\'{e}re's law.}
 The purpose of such definitions  
is to write the laws of EM with notations specific to the international unit system 
(SI). 
\par The law of electromagnetic induction is written as
$$\oint\limits_{\Gamma}\Evec\cdot\bbox{dl}
=-\frac{1}{\alpha_i}\intl_{\Sigma_\Gamma}\frac{\d\Bvec}
{\d t}\cdot\bbox{dS}\;\;\mbox{or}\;\;\nablav\times\Evec=-\frac{1}{\alpha_i}
\frac{\d\Bvec}
{\d t},\;\;\alpha_i>0\;$$
where $\alpha_i$ is the last constant introduced. Hence, the Maxwell's equations are 
\begin{equation}\label{maxw1}
\frac{1}{\tilde{\mu}_0}\nablav\times\Bvec=\frac{1}{\al_0}\jvec+\frac{\tilde{\eps}_0}
{\al_0}
\frac{\d\Evec}{\d t},
\end{equation}
\begin{equation}\label{maxw2}
\nablav\times\Evec=-\frac{1}{\alpha_i}\frac{\d\Bvec}{\d t},
\end{equation}
\begin{equation}\label{maxw3}
\nablav\cdot\Bvec=0,
\end{equation} 
\begin{equation}\label{maxw4}
\nablav\cdot\Evec=\frac{1}{\tilde{\eps}_0}\rho.
\end{equation}
Several arguments justify the equality between $\al_0$ and $\al_i$.
One may argue that 
$$\alpha_i=\al_0$$
as in Jackson's book \cite{jack}, but we may use directly the equation \eref
{FL} considering a closed conductor $\Gamma$ moving with a constant velocity 
$\bbox{u}$ in an external nonuniform                                                                                                                                                                                                                                                                                                                      magnetic field $\Bvec(\rvec)$ in the 
laboratory system $L$. The electromotive force corresponding to the 
 electric current due to the Lorentz force is \cite{tamm}$$
\mathcal{E}=\frac{1}{\al_0}\oint\limits_{\Gamma}\left(\bbox{u}\times\Bvec
\right)\cdot\bbox{dl}=\frac{1}{\al_0}\frac{1}{\delta t}\oint\limits_{\Gamma}
\left( \bbox{dl}\times\bbox{\delta a}\right)\cdot\Bvec=
-\frac{1}{\al_0}\frac{\delta N_m}{\delta t}$$
where $\bbox{\delta a}$ is the displacement vector of the element $\bbox{dl}$ 
of the current contour during the time $\delta t$, and $\delta N_m$ is the variation  
  of the magnetic flux $N_m$ through a surface attached to this contour due to the 
  displacement $\bbox{\delta a}$. 
   After comparing the 
last equation to the equation \eref{maxw2}, and after considering that only the magnetic flux 
variation is the defining element of the electromagnetic induction, we have to admit the 
equality between the two constants  $\al_i$ and $\al_0$.
\par In \cite{jack} this equality is argued theoretically using the Galilei invariance 
of Maxwell's equations for $u<<c$: for the observer in the system $L'$ of 
the conductor the effect is associated with an induced electric field
$$\Evec'=\frac{1}{\al_0}\bbox{u}\times\Bvec,\;\;(\Evec=0).$$
This equation, together with the transformation  law $\Bvec'=\Bvec$, is, indeed, the first 
approximation of the relativistic transformation law.
\par Another argument for considering the two constants equal is given  
  by the physical requirements of the EM theory. The definitions of charge, energy {\it etc} must complete the 
Maxwell's equations together with the corresponding theorems resulting from these 
equations. Combining the equations \eref{maxw1} and \eref{maxw2} one obtains 
$$\frac{1}{\tilde{\mu}_0}\Bvec\times\left(\nablav\times\Bvec\right)+
\tilde{\eps}_0\Evec\times\left(\nablav\times\Evec\right)=-\frac{1}{\al_0}\jvec\times
\Bvec-\frac{\tilde{\eps}_0}{\al_0}\frac{\d\Evec}{\d t}\times\Bvec-\frac{\tilde{\eps}_0}{
\al_i}\Evec\times\frac{\d\Bvec}{\d t}$$
and, finally, 
\begin{equation}\label{thimp}
\tilde{\eps}_0\left[\frac{1}{\al_0}\frac{\d\Evec}{\d t}\times\Bvec+\frac{1}{\al_i}
\Evec\times\frac{
\d\Bvec}{\d t}\right]=-\nablav\cdot\Tsf-\bbox{f}
\end{equation}
where  $\bbox{f}=\rho\Evec+(1/\al_0)\jvec\times\Bvec$ is the Lorentz force density,  
 and $${\sf T}_{ik}=\frac{1}{2}\left(\tilde{\eps}_0E^2+\frac{1}{\tilde{\mu}_0}B^2\right)
\delta_{ik}-\left(\tilde{\eps}_0E_iE_k+\frac{1}{\tilde{\mu}_0}B_iB_k\right).$$
The equation \eref{thimp} represents  the relation between the 
electromagnetic forces and the Maxwell's stress tensor ${\mathcal T}_{ik}=
-{\sf T}_{ik}$ within the static case. It can  be considered as the EM momentum theorem if the left hand 
term is a time derivative. Consequently, this requires $\al_i=\al_0$ and defines  
the electromagnetic momentum density by 
$$\bbox{g}_{em}=\frac{\tilde{\eps}_0}{\al_0}\left(\Evec\times\Bvec\right).$$
\par Let's consider  $\al_i=\al_0$  from now on. We point out that the equations 
 are written using  SI units by substituting $\al_0=1$ , $\tilde{\eps}_0=\eps_0, 
\;\tilde{\mu}_0=\mu_0$. \footnote{Those of my students who do not agree with my general notation 
are free to use this choice and, so, to work in SI units.  They are notified on this freedom
from the first class.}

\par In the case of the free  EMF ($\rho=0,\;\jvec=0$) one obtains the 
propagation equations for $\Evec$ and $\Bvec$,
$$\Delta\Evec-\frac{\tilde{\eps}_0\tilde{\mu}_0}{\al^2_0}\frac{\d^2\Evec}{\d t^2}=0,\;\;\;  
\Delta\Bvec-\frac{\tilde{\eps}_0\tilde{\mu}_0}{\al^2_0}\frac{\d^2\Bvec}{\d t^2}=0$$
From the last equations one obtains the fundamental relation 
\begin{equation}\label{fdm}
\;\;\;\;\;\;\;\;\;\;\;\;\;\;\;\;\;\;\;\;\;\;\;\;\;\;\;\;\;\;\;\;\;\;\;\;\;\;\;
\frac{\al^2_0}{\tilde{\eps}_0\tilde{\mu}_0}=c^2, \end{equation}
between the three constants $\tilde{\eps}_0$, $\tilde{\mu}_0$, $\al_0$ introduced in the 
Maxwell's equations and an experimental constant, the light speed $c$. The two remaining 
arbitrary constants define different unit systems. 
\par The electromagnetic potentials are introduced by the equations
$$\Bvec=\nablav\times\Avec,\;\;\;\Evec=\nablav\Phi-\frac{1}{\al_0}
\frac{\d\Avec}{\d t}$$
and verify
\beqan
\Delta\Avec&-&\frac{\tilde{\eps}_0\tilde{\mu}_0}{\al^2_0}\frac{\d^2\Avec}{\d t^2}=
-\frac{\tilde{\mu}_0}{\al_0}\jvec+\nablav.\left(\nablav\cdot\Avec+
\frac{\tilde{\eps}_0\tilde{\mu}_0}{\al_0}\frac{\d\Phi}{\d t}\right),\nonumber\\
\Delta\Phi&=&-\frac{1}{\tilde{\eps}_0}\rho-\frac{1}{\al_0}\frac{\d}{\d t}
\nablav\cdot\Avec.\eeqan
The gauge transformations are
$$\Avec\longrightarrow \Avec+\nablav\Psi,\;\;\;\Phi\longrightarrow 
\Phi-\frac{1}{\al_0}\frac{\d \Psi}{\d t},$$
and the Lorenz constraint is
$$\nablav\cdot\Avec+\frac{\tilde{\eps}_0\tilde{\mu}_0}{\al_0}\frac{\d\Phi}{\d t}=0.$$
Correspondingly, the equations of the potentials in this gauge are
$$\Delta\Avec-\frac{1}{c^2}\frac{\d^2\Avec}{\d t^2}=-\frac{\tilde{\mu}_0}{\al_0}\jvec,\;\;\;
\Delta\Phi-\frac{1}{c^2}\frac{\d^2\Phi}{\d t^2}=-\frac{1}{\tilde{\eps}_0}\rho$$
with the retarded solutions
\begin{equation}\label{ret}
\fl\Avec(\rvec,t)=\frac{\tilde{\mu}_0}{4\pi\al_0}\int\frac{\jvec(\rvec^{'},t-R/c)}
{R}\rmd^3x',\;\;\;\Phi(\rvec,t)=\frac{1}{4\pi\tilde{\eps}_0}\int\frac{\rho
(\rvec^{'},t-R/c)}{R}\rmd^3x'.\end{equation}

\par The above notations are specific for {\it rationalized } unit systems. To find out 
the changes required to rewrite the Maxwell's equations using {\it non rationalized} unit systems, 
let's consider the solutions \eref{ret} for the scalar and vector potentials. 
We convert the Maxwell's equations  \eref{maxw1}-\eref{maxw4} to {\it non rationalized} notations 
by eliminating the factor $1/4\pi$ from \eref{ret}
$$\tilde{\eps}_0\longrightarrow \frac{\tilde{\eps}_0}{4\pi},\;\;\;\tilde{\mu}_0\longrightarrow 
4\pi\tilde{\mu}_0.$$
The equation \eref{fdm} is invariant  to these transformations.
The  Maxwell's equations with the notations of a non rationalized 
unit system are 
$$\frac{1}{\tilde{\mu}_0}\nablav\times\Bvec=\frac{4\pi}{\al_0}\jvec+\frac{\tilde{\eps}_0}
{\al_0}\frac{\d\Evec}{\d t},$$
$$\nablav\times\Evec=-\frac{1}{\al_0}\frac{\d\Bvec}{\d t},$$
$$\nablav\cdot\Bvec=0,$$
$$\nablav\cdot\Evec=\frac{4\pi}{\tilde{\eps}_0}\rho.$$

\par The Maxwell's equations  
\eref{maxw1}-\eref{maxw4} with $\al_i=\al_0$  
 and all their consequences are rewritten   in all usual unit systems by substituting the following values  
 for the three constants:
\beqan
&~&\mbox{SI:}\;\;\;\;\;\;\;\;\;\;\;\;\;\;\;\al_0=1,\;\;\tilde{\eps}_0=\eps_0\;\;\;\;\;
\;,\tilde{\mu}_0=\mu_0\;\;\;,\\
&~& \mbox{Heaviside:}\;\;\; \al_0=c,\;\;
\tilde{\eps}_0=\tilde{\mu}_0=1,\\
&~&\mbox{Gauss:}\;\;\;\;\;\;\;\;
\al_0=c,\;\;\tilde{\eps}_0=\frac{1}{4\pi},\;\;\;\;\tilde{\mu}_0=4\pi,\\
&~&\mbox{esu:}\;\;\;\;\;\;\;\;\;
\;\;\;\al_0=1,\;\;\tilde{\eps}_0=\frac{1}{4\pi},
\;\;\;\;\tilde{\mu}_0=\frac{4\pi}{c^2},\\
&~&\mbox{emu:}\;\;\;\;\;\;\;\;\;\;\al_0=1,\;\;\tilde{\eps}_0=\frac{1}{4\pi c^2},
\;\;\tilde{\mu}_0=1.
\eeqan

\par These notations work also within the relativistic electrodynamics (in vacuum). 
The  relativistic equations of motion of a charged particle are obtained from 
the Lagrange function 
$$L(t)=-m_0c^2\sqrt{1-\frac{v^2}{c^2}}-q\Phi+\frac{1}{\al_0}q\bbox{v}
\cdot\Avec$$
which may be written with an invariant parameterization $\lambda$ as
$$L(\lambda)=-m_0c\sqrt{\dot{x}_\mu(\lambda)\dot{x}^\mu(\lambda)}
-\frac{1}{\al_0}A_\mu(x)\dot{x}^\mu$$
where  
$$\left(A^\mu\right)=\left(\frac{\al_0}{c}\Phi,\;\Avec\right).$$  is the {\it 4-potential}.
The relativistic invariance of the motion equations  and of the Maxwell's 
equations is realized by defining $T_{\mu\nu}=\d_\mu A_\nu-\d_\nu A_\mu$ as 
 components of a tensor. In particular, $A^\mu$ may be considered as 
the components of a 4-vector although this is not a necessary condition 
of the invariance of the theory, except the request to have covariant 
equations for the 4-potential.
\footnote{Curiously, in a considerable part of the physics literature, the vectorial character 
of the 4-potential is presented as a necessary condition for the relativistic 
invariance of the theory.} 
\par As it is pointed out in \cite{jack,leung},  many difficulties are 
encountered trying to generalize this procedure to the macroscopic 
electromagnetic field in the presence of a medium. The complications arise 
due to some inconsistencies in the definitions adopted in various unit 
systems as it is pointed out in \cite{leung}.\\
In \cite{vre93} one defines the vectors $\bbox{D}$ and $\bbox{H}$ by the 
equations
$$\bbox{D}=\tilde{\eps}_0\Evec+\bbox{P},\;\;\;\bbox{H}=
\frac{1}{\tilde{\mu}_0}\Bvec-\bbox{M}$$
such that one may obtain the macroscopic Maxwell's equations only in SI and Heaviside 
unit systems. In \cite{leung},  the definitions of $\bbox{D}$ and $\bbox{H}$ 
are given by introducing two new constants $\al_d$ and $\al_h$ (in \cite{leung} labeled  $k_D$  
and $k_H$). With our notations, they are  
$$\bbox{D}=\al_d\left(\Evec+\frac{1}{\tilde{\eps}_0}\bbox{P}\right),$$
$$\bbox{H}=\al_h\left(\Bvec-\tilde{\mu}_0\bbox{M}\right).$$
However, in \cite{leung} a new constant $\al_m$ (in \cite{leung} labeled $k_M$) is introduced 
by the relation between the magnetization current $\jvec_m$ and the 
magnetization vector $\bbox{M}$
$$\jvec_m=\al_m\nablav\times\bbox{M}.$$
This new constant is not necessary. Actually, we can reduce the number of supplementary 
 constants to two, as in \cite{jack}.

 The constant $\al_m$ is necessary in the definition of the magnetic 
dipolar moment (and in all multipolar orders). If we consider the Laplace's 
force, the corresponding expression is well defined by the Lorentz force.  
Furthemore, for a steady current loop $\Gamma$ we have
$$\bbox{F}(I,\Gamma)=\frac{I}{\al_0}\oint\limits_\Gamma \bbox{dl}\times\Bvec$$
We may demonstrate the relation 
$$\oint\limits_\Gamma \bbox{dl}\times\Bvec
= \intl_{\Sigma(\Gamma)}(\bbox{n}\cdot\nablav)\Bvec
\rmd S.$$
So, the Laplace's force is equivalent, at least regarding the resulting force, 
to a fictitious force acting on the shell $\Sigma(\Gamma)$
$$\bbox{dF}^{\,'}=\frac{I}{\al_0}\bbox{dS}.$$
The  simplest and natural definition of the magnetic moment $\bbox{dm}$ 
corresponding to an elementary shell is
$$\bbox{dm}=\frac{I}{\al_0}\bbox{dS}.$$
Also, it is possible  to relate the torque $\bbox{dm}\times\Bvec $ to the 
Laplace's force.\cite{vre93}. Therefore, in the case of a current distribution 
$\jvec$ in ${\mathcal D}$ the magnetic dipolar moment is defined by
$$\bbox{m}=\frac{1}{2\al_0}\intl_{\mathcal D}\rvec\times\jvec\rmd^3x.$$
Generally, one may define the $n-$th order magnetic moment by the tensor 
\cite{castel} 
\begin{equation}\label{cast}
\msf^{(n)}=\frac{n}{(n+1)\al_0}\intl_{\mathcal D}\rvec^n\times\jvec\;\rmd^3x.
\end{equation}
The magnetization current $\jvec_m$ is given by the relation
$$\jvec_m=\al_0\,\nablav\times\bbox{M}$$
where $\bbox{M}$ includes the contributions of all magnetic multipoles. 
This is a result of the definition \eref{cast} and of the average of 
microscopic equations of EMF. In conclusion, the equality
$$\al_m=\al_0$$
is justified while a third supplementary constant 
 is not necessary in the case of the macroscopic field.

\section{Conclusion}
By writing the macroscopic Maxwell's equations as in \cite{vre93},
\beqan
\nablav\times\bbox{H}&=&\frac{1}{\al_0}\jvec+\frac{1}{\al_0}\frac{\d\bbox{D}}
{\d t},\\
\nablav\times\Evec&=&-\frac{1}{\al_0}\frac{\d\Bvec}{\d t},\\
\nablav\cdot\Bvec&=&0,\\
\nablav\cdot\bbox{D}=\rho\eeqan
only the equations in SI and Heaviside's systems are obtained. To change the unit system  
from Heaviside's  to the Gaussian one, we have to memorize some factors 
 $4\pi$.
\par With notations from the present paper we have
\beqa\label{Mmacro}
\nablav\times\bbox{H}&=&\frac{\tilde{\mu}_0\al_h}{\al_0}\jvec+\frac{\al_0\al_h}
{\al_d}\frac{1}{c^2}\frac{\d\bbox{D}}{\d t},\nonumber\\
\nablav\times\Evec&=&-\frac{1}{\al_0}\frac{\d\Bvec}{\d t},\nonumber\\
\nablav\cdot\Bvec&=&0,\nonumber\\
\nablav\cdot\bbox{D}&=&\frac{\al_d}{\tilde{\eps}_0}\rho,\eeqa
and
\begin{equation}\label{DH}
\bbox{D}=\al_d\left(\Evec+\frac{1}{\tilde{\eps}_0}\bbox{P}\right),\;\;
\bbox{H}=\al_h\left(\Bvec-\tilde{\mu}_0\bbox{M}\right).\end{equation}
The equations \eref{Mmacro} result from the microscopic equations using 
the relations 
$$<\rho_{\scriptstyle micro}>=-\nablav\cdot\bbox{P},\;\;\;
<\jvec_{\scriptstyle micro}>=\jvec +\al_0\nablav\times \bbox{M}+
\frac{\d\bbox{P}}{\d t},$$
and the definitions \eref{DH}.
The following values for the two supplementary constants, named in \cite{leung} 
{\it conventional} constants,  should be substituted 
to obtain the equations within different unit systems.
 For the two supplementary constants in the usual unit we have the 
following expressions:
\beqan
\mbox{SI:}\;\;\;\;\;\;\;\;\;\;\;\;\;\;\;\;\;\;\;\al_d&=&\eps_0,
\;\;\;\;\;\;\;\;\al_h=\frac{1}{\mu_0},\\
\mbox{Heaviside}:\;\;\;\;\;\al_d&=&1,\;\;\;\;\;\;\;\;\;\al_h=1,\\
\mbox{Gaussian}:\;\;\;\;\;\;\al_d&=&1,\;\;\;\;\;\;\;\;\;\al_h=1,\\
\mbox{esu}:\;\;\;\;\;\;\;\;\;\;\;\;\;\;\al_d&=&1,\;\;\;\;\;\;\;\;\;\al_h=c^2,\\
\mbox{emu}:\;\;\;\;\;\;\;\;\;\;\;\;\;\al_d&=&\frac{1}{c^2},\;\;\;\;\;\;\;\al_h=1
\eeqan
Although the number of the {\it conventional} constants is reduced to two, the 
conclusion from \cite{leung} remains valid: the complications due to these 
{\it conventional} constants make the result "not as appealing as that obtained 
in the vacuum case...".
\vspace{1cm}
\par
{\bf Acknowledgments}\\

\par I thank Dr Sorina Zota for reading the manuscript and for many helpful comments.

\vspace{1.cm}
\par 


{\bf References}
\begin{thebibliography}{10}
\bibitem{leung}
Leung P T 2004 {\it A note on the 'system-free' expressions of Maxwell's 
equations}, {\it Eur. J. Phys.} {\bf 25} 
~~~~~~~~~~(online at stacks.iop.org/EJP/25/N1)
\bibitem{jack}
Jackson J D 1975 {\it Classical Electrodynamics} (Wiley New York) p.755-8
\bibitem{vre93}
Vrejoiu C  1993  {\it Electrodynamics and Relativity Theory }(in romanian)
EDP
\bibitem{tamm}
Tamm I E 1979 {\it Fundamentals of the Theory of Electricity} (Mir Publishers Moscow)
\bibitem{castel}
Castellanos A, Panizo M, Rivas J 1978 {\it Am.J.Phys.}, {\bf 46} 1116-17

\end{thebibliography}
\end{document}